# Analysis of Biometric Authentication Protocols in the Blackbox Model

Koen Simoens, Julien Bringer, Hervé Chabanne, and Stefaan Seys

January 3, 2011

*Abstract*—In this paper we analyze different biometric authentication protocols considering an internal adversary. Our contribution takes place at two levels. On the one hand, we introduce a new comprehensive framework that encompasses the various schemes we want to look at. On the other hand, we exhibit actual attacks on recent schemes such as those introduced at ACISP 2007, ACISP 2008, and SPIE 2010, and some others. We follow a blackbox approach in which we consider components that perform operations on the biometric data they contain and where only the input/output behavior of these components is analyzed.

*Keywords*—Biometrics, template protection, authentication, protocols, blackbox security model, malicious adversaries

## I. INTRODUCTION

ALTHOUGH biometric template protection is a relatively young discipline, already over a decade of research has brought many proposals. Methods to secure biometric data can be separated in three levels. The first one is to have biometric data coming in a self-protected form. Many algorithms have been proposed: quantization schemes [1], [2] for continuous biometrics; fuzzy extractors [3] and other fuzzy schemes [4]–[6] for discrete biometrics; and cancellable biometrics [7]–[9]. The security of such template-level protection has been intensively analyzed, e.g., in [10]–[13]. On a second level one can use hardware to obtain secure systems, e.g., [14], [15]. Finally, at a third level advanced protocols can be developed to achieve biometric authentication protocols relying on advanced cryptographic techniques such as Secure Multiparty Computation, homomorphic encryption or Private Information Retrieval protocols [16, Ch. 9] [17]–[24].

The focus of our work is on this third level. In this work, we analyze and attack some existing biometric authentication protocols. We follow [25] where an attack against a hardware-assisted secure architecture [15] is described. The work of [25] introduces a blackbox model that is taken back and extended here. In this blackbox model, internal adversaries are considered. These adversaries can interact with the system by using available input/output of the different functionalities. Moreover, the adversaries are malicious in the sense that they can deviate from the honest-but-curious classical behaviour, which is most often assumed.

Our contributions are the following. We extend the blackbox framework initiated in [25] with the distributed system model of [19] in a way that it can handle different existing proposals for biometric authentication. We show how this blackbox approach can lead to attacks against these proposals. We describe in detail our analysis of three existing protocols [19], [20], [22] and give arguments on some others [23], [24]. In the framework we propose, we study how the previous attacks can be formalized. We list all the possible existing attacks points and the different internal entities that can lead the attacks, and we reveal the potential consequences.

The rest of the paper is organized as follows. The framework is developed in Section II and introduces the system and attack model. This is then applied to existing protocols in Section III, where detailed attacks are described. Section IV formalizes these attacks and Section V concludes the paper.

## II. FRAMEWORK

In this section we present a framework that forms a basis for the security analysis of biometric authentication protocols. The framework models a generic distributed biometric system and the (internal) adversaries against such system. We define the roles of the different entities that are involved and their potential attack goals. From these roles and attack goals we derive the requirements that are imposed on the data that are exchanged between the entities.

*Biometric Notation:* Two measurements of the same biometric characteristic are never exactly the same. Because of this behavior, a biometric characteristic is modeled as a random variable $B$, with distribution $p_B$ over some range $\mathcal{B}$. A sample is denoted as $b$. Two samples or templates are related if they originate from the same characteristic. In practice, we will say they are related if their mutual distance is less than some threshold. Therefore, a distance function d is defined over $\mathcal{B}$ and for each value in the range of d that is used as the threshold when comparing two samples a false match rate (FMR) and a false non-match rate (FNMR) can be derived.

Biometric variables can be continuous or discrete but in the remainder of the paper we will assume that they are discrete. Note that the variables may consist of multiple components. For example, a sample may consist of a bitstring, which is the quantized version of a feature vector, and an other bitstring that indicates erasures or unreliable components in the first and thus act as a mask.

### A. System Model

Our system model follows to a large extent the model defined by Bringer *et al.* [19], which was also used to define new schemes in [20] and [26]. This model is motivated by

4a separation-of-duties principle: the different roles for data processing or data storage on a server are separated into three distinct entities. Using distributed entities is a baseline to avoid one to control all information and it is a realistic representation of how current biometric systems work in practice (cf. [27]).

*System Entities:* The different entities involved in the system are a user $\mathcal{U}_i$, a sensor $\mathcal{S}$, an authentication server $\mathcal{AS}$, a database $\mathcal{DB}$ and a matcher $\mathcal{M}$. User $\mathcal{U}_i$ wishes to authenticate to a particular service and has, therefore, registered his biometric data $b_i$ during the *enrollment* procedure. In the context of the service the user has been assigned an identifier $ID_i$, which only has meaning within this context. The biometric reference data $b_i$ are stored by $\mathcal{DB}$, who links the data to identifier $i$. The mapping from $ID_i$ to $i$ is only known by $\mathcal{AS}$, if relevant. Note that in some applications it is possible that the same user is registered for the same service or in the same database with different samples, $b_i$ and $b_j$, and different identities, i.e., $ID_i \neq ID_j$ in the service context or $i \neq j$ in the database context. The property of not being able to relate queries under these different identities is the *identity privacy* requirement as defined in [19].

During the *authentication* procedure the sensor $\mathcal{S}$ captures a fresh biometric sample $b_i'$ from user $\mathcal{U}_i$ and forwards the sample to $\mathcal{AS}$. The authentication server $\mathcal{AS}$ manages authorizations and controls access to the service. To make the authorization decision, $\mathcal{AS}$ will rely on the result of the biometric *verification* or *identification* procedure that is carried out by the matcher $\mathcal{M}$. It is assumed that there is no direct link between $\mathcal{M}$ and $\mathcal{DB}$. As such, $\mathcal{AS}$ requests from $\mathcal{DB}$ the reference data that are needed by $\mathcal{M}$ and forwards them to $\mathcal{M}$. It is further assumed that the system accepts only biometric credentials. This means that the user provides his biometric data and possibly his identity, but no user-specific key, password or token. Fig. 1 shows the participating entities.

*Functional Requirements:* Enrollment often involves offline procedures, like identity checks, and is typically carried out under supervision of a security officer. Therefore, we assume that users are enrolled properly and only authentication procedures are analyzed in our framework. A distinction has to be made between verification and identification. Verification introduces a *selection step*, which implies that $\mathcal{DB}$ returns only one of its references, namely the $b_i$ that corresponds to the identifier $i$ that is used in the context of the database. The entity that does the mapping between $ID_i$ and $i$, when applicable, is generally $\mathcal{AS}$. In identification mode, $\mathcal{DB}$ returns the entire set of references, in some protected form, to $\mathcal{AS}$. The database can then be combined with $b_i'$ and forwarded to $\mathcal{M}$. The matcher $\mathcal{M}$ has to verify that $b_i'$ matches with one or a limited number of $b_i$ in the received set of references or that one of the matching references has index $i$.

We define the minimal logical functionality to be provided by our system entities in terms of generic information flows, which are included in our model in Fig. 1. In this functional model, we represent the result of the biometric comparison as a function of the distance $d(b_i', b_i)$. This is a generic representation of the actual comparison method: $\mathcal{M}$ can evaluate simple distances but also run more complex comparisons and will output either similarity measures or decisions that are based on some threshold $t$. The information flows are as follows.

User $\mathcal{U}_i$ presents a biometric characteristic $B_i$ that will be sampled by the sensor $\mathcal{S}$ to produce a sample $b_i'$. When operating in verification mode $\mathcal{U}_i$ will claim an identity $ID_i$:

$$\mathcal{U}_i \xrightarrow{b_i' \leftarrow B_i} \mathcal{S} \qquad \text{or} \qquad \mathcal{U}_i \xrightarrow{b_i' \leftarrow B_i,\, ID_i} \mathcal{S}. \qquad (1)$$

The sensor $\mathcal{S}$ forwards $b_i'$ and $ID_i$ in some form to $\mathcal{AS}$:

$$\mathcal{S} \xrightarrow{f_1(b_i')} \mathcal{AS} \qquad \text{or} \qquad \mathcal{S} \xrightarrow{f_1(b_i'),\, g_1(ID_i)} \mathcal{AS}. \qquad (2)$$

In general $g_1(ID_i) = ID_i$ but it can also be a mapping to an encrypted value to hide $ID_i$ from $\mathcal{AS}$. If applicable $\mathcal{AS}$ resolves the mapping $g_1(ID_i)$ to the identifier $i$ and requests reference data for one or more users from $\mathcal{DB}$ by sending at least one request $g_2(b_i', i)$:

$$\mathcal{AS} \xrightarrow{g_2(b_i', i)} \mathcal{DB}. \qquad (3)$$

Note that the function $g_2$ does not necessarily use all the information in its arguments, e.g., the fresh sample $b_i'$ may be ignored.

Database $\mathcal{DB}$ provides $\mathcal{AS}$ with reference data for one or more users in some form. It is possible that $\mathcal{DB}$ returns the entire database, e.g., in case of identification:

$$\mathcal{AS} \xleftarrow{f_2(\{b_i\})} \mathcal{DB}. \qquad (4)$$

The authentication server $\mathcal{AS}$ forwards the fresh sample $b_i'$ and the reference data $b_i$ in some combined form to $\mathcal{M}$:

$$\mathcal{AS} \xrightarrow{f_3(b_i', \{b_i\})} \mathcal{M}. \qquad (5)$$

Note that $\mathcal{AS}$ has only $f_1(b_i')$ and $f_2(b_i)$ at his disposal to compute $f_3(b_i', \{b_i\})$.

The matcher $\mathcal{M}$ performs a biometric comparison procedure on the received $b_i'$ and $\{b_i\}$ and returns the result to $\mathcal{AS}$. The result may contain decisions or scores or different identities but should at least be based on one distance calculation between the fresh sample $b_i'$ and a reference $b_i$:

$$\mathcal{AS} \xleftarrow{f_4(\mathrm{d}(b_i', \{b_i\}))} \mathcal{M}. \qquad (6)$$

Different data are stored by the different entities. The database stores references $\{b_i\}$. The authentication service stores the information needed to map $g_1(ID_i)$ to $i$, if applicable. The matchers can store non-biometric verification data, e.g., hashes of keys extracted from biometrics, or decryption keys that are use to recover the result of combining sample and reference. Also, the sensor can store key material to encrypt the fresh sample.

### B. Adversary Model

*Attacker Classification:* Based on the physical entry point of an attack a distinction is made between two types of attackers: *internal* attackers are corrupted components in the system and *external* attackers are entities that only have access to a communication channel. We will consider here only the issue of an insider attacker. As a baseline, we make the following assumption.

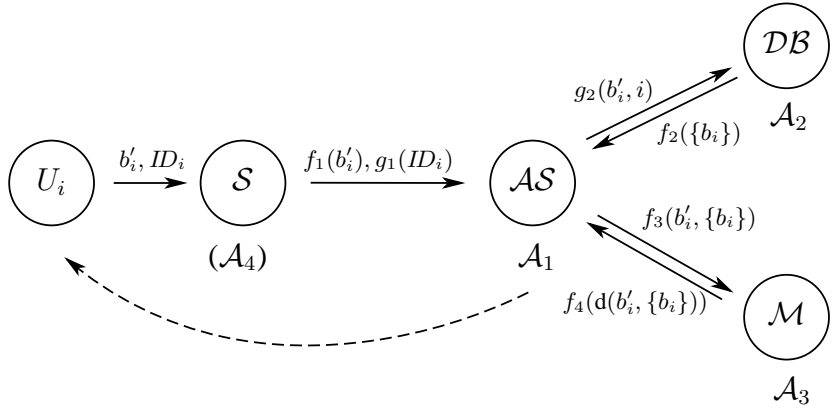

Fig. 1. System model with indication of generic information flows and attack points $\mathcal{A}_i$. User $\mathcal{U}_i$'s biometric is sampled by sensor $\mathcal{S}$. The sample $b'_i$ and $\mathcal{U}_i$'s identity are forwarded to the authentication server $\mathcal{AS}$, who requests the corresponding reference $b_i$ from database $\mathcal{DB}$. $\mathcal{AS}$ combines the sample and the reference and forwards the result to matcher $\mathcal{M}$, who performs the actual comparison and returns the result to $\mathcal{AS}$. The solid arrows represent the messages exchanged between the system entities. The dashed arrow represents the implicit feedback on the authentication result to the user $\mathcal{U}_i$, i.e., access to the requested service is granted if the sample matches the reference.

*Assumption 1:* The protocol ensures the security of the scheme against any external attacker.

As this can be reached by classical secure channel techniques, by an external security layer independent of the core protocol specification, we study further only the internal layer. Note that the security of the scheme needs to be expressed in terms of specific attack goals, which will be defined in the next section.

A second distinction is made based on an attacker's capabilities. Passive attackers or honest-but-curious attackers are attackers that only eavesdrop the communications in which they are involved and that can only observe the data that passes through them. They always follow the protocol specifications, never change messages and never generate additional communication. Active or malicious attackers are internal components that can also modify existing or real transactions passing through them and that can generate additional messages. We mainly focus on malicious internal attackers and we formulate the following additional assumption.

*Assumption 2:* The protocol ensures the security of the scheme against honest-but-curious entities, i.e. internal system components that always follow the protocol specifications but eavesdrop internal communication.

We will explain in Section II-C how this has a direct impact on the properties of the different functionalities of our model.

Finally, we put aside the threats on the user or client side, by concentrating the analysis on the remote server's side, i.e., $\mathcal{AS}$, $\mathcal{DB}$ or $\mathcal{M}$. The information leakage for the user and the client is generally only the authentication or identification result. They can, however, try to gain knowledge on the reference data $b_i$ by running queries with different $b'_i$, e.g., in some kind of hill climbing attack. The difficulty can highly vary depending on the modalities, the threshold and the scenario. A basic line of defense is to limit the number of requests, to ensure the aliveness of the biometric inputs provided by the user and to hide the result when applicable. Although it is important to implement such defense mechanisms, the threats are inherent to any biometric authentication or identification system. So we do not take the user or the sensor into account as an attacker in this model and the primary attack points are $\mathcal{AS}$, $\mathcal{DB}$ and $\mathcal{M}$. Nonetheless, there may be inside attackers that also control the biometric inputs to some extent. We model this with a secondary attack point at the sensor.

*Assumption 3:* The user $\mathcal{U}_i$ or the sensor $\mathcal{S}$ cannot be attackers on their own but they can act as a secondary attack point in combination with a primary attack point at $\mathcal{AS}$, $\mathcal{DB}$ or $\mathcal{M}$. If this the case an attacker can choose the input sample $b'_i$ through $\mathcal{S}$ and observe whether the authentication request was successful through $\mathcal{U}_i$.

Of course, the baseline assumptions have to be checked before proceeding with a full analysis of the security of a scheme, but as such, they clarify what the big issues are that may remain in state-of-the-art schemes. They also underline what the hardest challenges are when designing a secure biometric authentication protocol. Fig. 1 sums up the different attack points we consider in our attack model.

*Attack Goals:* As noted above, the security of a scheme is expressed in terms of specific attack goals or adversary objectives. Therefore, we define the following global attack goals.

- **Learn reference** $b_i$ **.** In accordance to the security definitions in [25] we define different gradations in the information that an attacker may want to learn from $b_i$. Minimum leakage refers to the minimum information that allows, e.g., linking of references with high probability. Authorization leakage is the information that is needed to construct a sample that is within distance $t$, the system threshold, of the reference $b_i$. Full leakage gives full knowledge of $b_i$. When a scheme is resistant to this attack in all three gradations we say that it provides *biometric reference privacy*.
- **Learn sample** $b'_i$ **.** The same gradations apply as in the previous attack goal. We call the security property associated with this attack goal *biometric sample privacy*.
- **Trace users with different identities.** This attack can be achieved when different references from the same user, possibly coming from different applications, can be linked. A system that is resistant to such attack is said to provide *identity privacy* [26].

TABLE I
RELEVANCE OF ATTACK GOALS FOR DIFFERENT (MALICIOUS) ENTITIES IN THE SYSTEM MODEL (? = ONLY RELEVANT IF THE SCHEME UNDER CONSIDERATION WAS DESIGNED TO HIDE REFERENCES FROM $\mathcal{DB}$; * = ONLY RELEVANT IF THE PROTOCOL OPERATES IN IDENTIFICATION MODE OR IF $ID_i$ AND $i$ ARE HIDDEN FROM $\mathcal{AS}$ IN VERIFICATION MODE).

| Attack goal | $\mathcal{AS}$ | $\mathcal{DB}$ | $\mathcal{M}$ |
|---|---|---|---|
| Learn $b_i$ | V | ? | V |
| Learn $b'_i$ | V | V | V |
| Trace $\mathcal{U}_i$ with different identities | V | ? | V |
| Trace $\mathcal{U}_i$ over different queries | V* | V | V |

- **Trace users over different queries.** This attack refers to linking queries, whether anonymized or not, based on $i$, $b_i$ or $b'_i$. The property of a system that prevents such attack is called *transaction anonymity* [26]. Note that an attacker that is able to learn $b'_i$ can automatically trace users based on the learned sample.

The formulated attack goals may apply to the different internal attackers as indicated by the different attack points. The relevance of the attack goals is shown in TABLE I. Attack goals can be generalized for combinations of inside attackers, e.g., $\mathcal{AS}$ and $\mathcal{M}$. They are relevant for the combination if they are relevant for each attacker individually. As a counterexample, learning $b_i$ is not always relevant for the combination $\mathcal{AS}$-$\mathcal{DB}$. In some schemes it is assumed that $\mathcal{DB}$ stores references in the clear so the attack "learn $b_i$" becomes trivial. It is important, however, that such schemes explicitly mention the assumption that $\mathcal{DB}$ is fully trusted. It will become clear in the further sections that the main focus of our work is on $\mathcal{AS}$ who is a powerful attacker. This way of thinking is rather new and many protocols are not designed to be resistant to such attacker.

For each attacker or combination of attackers, and for each relevant attack goal a security requirement can be defined, namely that the average success probability of the given attacker that mounts the given attack on the scheme should be negligible in terms of some security parameter defined by the application. When analyzing the security of biometric authentication protocols that include distributed entities, each of these requirements should be checked individually.

### C. Requirements on Data Flows

Coming back to the functionalities in our system model (cf. Section II-A), we use the attack goals defined in TABLE I to impose requirements on the data that are being exchanged.

- $\mathcal{AS}$ should not be able to learn $b'_i$ hence $f_1$ is at least one-way, meaning that $b'_i$ should be unrecoverable from $f_1(b'_i)$ with overwhelming probability. To prevent tracing $\mathcal{U}_i$ over different queries, e.g., in identification mode, it could also be required that $f_1$ is semantically secure. We note that semantic security is a security notion that might be too strong but it ensures that the function prevents the minimum leakage as described under attack goal *learn $b_i$* (Section II-B).

- $\mathcal{AS}$ should not learn $b_i$ hence $f_2$ is at least one-way. To prevent tracing users with different identities it may be required that $f_2$ is also semantically secure.
- If applicable, $\mathcal{AS}$ should not be able to trace $\mathcal{U}_i$ by linking queries on $ID_i$ or $i$, and thus $g_1$ should be semantically secure.
- If applicable, $\mathcal{DB}$ may not learn $b_i$, hence the $b_i$ would need to be stored in protected form using some semantically secure function.
- $\mathcal{DB}$ may not learn $b'_i$, hence $g_2$ is one-way on its first input. It should also be semantically secure to prevent tracing $\mathcal{U}_i$.
- $\mathcal{DB}$ may not be able to link the queries at all, hence $g_2$ should also be semantically secure on its second input.
- $\mathcal{M}$ may not learn the individual $b_i$ or $b'_i$ and must not be able to link references or queries from the same $\mathcal{U}_i$, hence $f_3$ should be semantically secure on tuples $\langle b'_i, b_j \rangle$.

Now as we demand that $\mathcal{M}$ returns a result to $\mathcal{AS}$ that is a function ($f_4$) of the distance $d(b'_i, b_i)$ while maintaining the confidentiality and the privacy of the data, this means that some operations must be malleable. Malleability refers to the property of some cryptosystems that an attacker can modify a ciphertext into another valid ciphertext that is the encryption of some function of the original message, but without the attacker knowing this message. Depending on the exact step when the combination of $b_i$ and $b'_i$ is realized, either $g_2$, $f_2$ or $f_3$ would be malleable. In the following section, we will show the impact of this fundamental limitation and how it can be exploited to attack existing protocols.

## III. APPLICATION TO EXISTING CONSTRUCTIONS

In this section, we begin to extend attacks that have been introduced by Bringer *et al.* in [25] in the context of hardware security to more complex cryptographic protocols that use homomorphic encryption in Section III-A for a scheme by Bringer *et al.* [19] and in Section III-B for a scheme by Barbosa *et al.* [20]. We then describe another kind of attacks by looking at a scheme by Stoianov [22] in Section III-C. Finally, we briefly discuss attacks on two other schemes [23], [24] in Section III-D. All schemes are described with the goal to fit them directly into our model.

### A. Bringer et al. ACISP 2007

*1) Description:* In [19], Bringer *et al.* presented a new security model for biometric authentication protocols that separates the tasks of comparing, storing and authorizing an authentication request amongst different entities: a fully trusted sensor $\mathcal{S}$, an authentication server $\mathcal{AS}$, a database $\mathcal{DB}$ and a matching service $\mathcal{M}$. The goal was to prevent any of the latter three to learn the relation between some identity and the biometric features that relate to it. Their model forms the basis of our current framework and in this model they presented a scheme that applies the Goldwasser-Micali cryptosystem [28]. Let $\mathcal{E}_{\text{GM}}$ and $\mathcal{D}_{\text{GM}}$ denote encryption and decryption, respectively, and note that for any $m, m' \in \{0, 1\}$ we have the homomorphic property $\mathcal{D}_{\text{GM}}(\mathcal{E}_{\text{GM}}(m, pk) \times \mathcal{E}_{\text{GM}}(m', pk), sk) = m \oplus m'$. The scheme in [19] goes as follows.

During the enrollment phase, the user $\mathcal{U}_i$ registers at the authentication server $\mathcal{AS}$. He then gets an index $i$ and a pseudonym $ID_i$. Let $N$ denote the total number of records in the system. Database $\mathcal{DB}$ receives and stores $(b_i, i)$ where $b_i$ stands for $\mathcal{U}_i$'s biometric template, a binary vector of dimension $M$, i.e., $b_i = (b_{i,1}, b_{i,2}, \ldots, b_{i,M})$. In the following, we suppose that $i$ is also the index of the record $b_i$ in the database $\mathcal{DB}$.

A key pair is generated for the system. Matcher $\mathcal{M}$ possesses the secret key $sk$. The public key $pk$ is known by $\mathcal{S}$, $\mathcal{AS}$ and $\mathcal{DB}$. The authentication server $\mathcal{AS}$ stores a table of relations $(ID_i, i)$ for $i \in \{1, \ldots, N\}$. Database $\mathcal{DB}$ contains the enrolled biometric data $b_1, \ldots, b_N$

When user $\mathcal{U}_i$ wants to authenticate himself, the $\mathcal{S}$ will send an encrypted sample $\mathcal{E}_{\text{GM}}(b'_i, pk)$ and $ID_i$ to $\mathcal{AS}$. The authentication server $\mathcal{AS}$ will request the encrypted reference $\mathcal{E}_{\text{GM}}(b_i, pk)$ from $\mathcal{DB}$ and combine it with the encrypted sample. Because of the homomorphic property, $\mathcal{AS}$ is able to obtain $\mathcal{E}_{\text{GM}}(b'_i \oplus b_i, pk)$. Note that the encryption is bitwise so $\mathcal{AS}$ will permute the $M$ encryptions and forward these to $\mathcal{M}$. Because $\mathcal{M}$ has the secret key $sk$, $\mathcal{M}$ can decrypt the permuted XOR-ed bits and compute the Hamming distance between the sample and the reference.

The security of this protocol is proved in [19] under the assumption that all the entities in the system will not collude and are honest-but-curious. It is this assumption that we challenge in our framework, which leads to the following attack.

*2) Authentication Server Adversary ($\mathcal{A}=\mathcal{AS}$):* The following attack shows how a malicious authentication server $\mathcal{AS}$ can learn the enrolled biometric template $b_i$ corresponding to some identity $ID_i$. To do so the authentication server $\mathcal{AS}$ requests the template $b_i$ without revealing $ID_i$ and receives from $\mathcal{DB}$ the encrypted template that was stored during enrolment, i.e., $\mathcal{E}_{\text{GM}}(b_i, pk) = \langle \mathcal{E}_{\text{GM}}(b_{i,1}, pk), \ldots, \mathcal{E}_{\text{GM}}(b_{i,M}, pk) \rangle$.

The attack consists of a bitwise search performed by $\mathcal{AS}$ in the encrypted domain. First $\mathcal{AS}$ computes the encryption of a zero bit $\mathcal{E}_{\text{GM}}(0, pk)$. If the public key is not known by $\mathcal{AS}$, he can take an encrypted bit of the template retrieved from $\mathcal{DB}$ and compute $\mathcal{E}_{\text{GM}}(b_{i,k}, pk)^0 = \mathcal{E}_{\text{GM}}(0, pk)$. Let the maximum allowed Hamming distance be $t$.

Now $\mathcal{AS}$ will take the first encrypted bit $\mathcal{E}_{\text{GM}}(b_{i,1}, pk)$, repeat it $t+1$ times and add $M-t-1$ encryptions of a zero bit. Note that the ciphertext $\mathcal{E}_{\text{GM}}(b_{i,1}, pk)$ can be re-randomized so that it is impossible to detect that the duplicate ciphertexts are "copies". If $b_{i,1}$ is one, the total Hamming distance as computed by $\mathcal{M}$ will be $t+1$ and $\mathcal{M}$ will return NOK (not ok). If $b_{i,1}$ is zero, the $\mathcal{M}$ will return OK. This process can be repeated for all bits of $b_i$, hence, $\mathcal{AS}$ can learn $b_i$ bit by bit in $M$ queries. To further disguise the attack $\mathcal{AS}$ can apply permutations and add up to $t$ encryptions of one-bits to make the query look genuine.

*3) Matcher and Sensor Adversary ($\mathcal{A}=\mathcal{M}+\mathcal{S}$):* A bitwise search attack similar to the previous attack can also be considered in the case of an adversary made of the matcher assisted by the sensor. The attack consists of the following steps:

- $\mathcal{S}$ sends the encryption of $\bar{0} = \langle 0, \ldots, 0 \rangle$;
- $\mathcal{M}$ receives $b_i \oplus \bar{0}$ bitwise but permuted and records the weight of $b_i \oplus \bar{0}$;
- $\mathcal{S}$ toggles a bit in the $\bar{0}$ vector in position $x$ and sends it to $\mathcal{AS}$;
- $\mathcal{M}$ observes the changed weight (+1 or -1) and learns the bit at position $x$ in $b_i$.

The adversary learns $b_i$ in $M$ queries.

*4) Discussion:* What makes the first attack ($\mathcal{A}=\mathcal{AS}$) feasible is that all bits are encrypted separately and that the cryptosystem is homomorphic and thus $f_1(b'_i)$ and $f_2(b_i)$ are malleable (needed to create the encryption of a zero-bit if the public key is not known). Moreover, it is not enforced that $\mathcal{AS}$ combines the input from the sensor and from the database.

To counteract this threat, one could require $\mathcal{S}$ to sign the input and force $\mathcal{DB}$ to merge the input with references, in this way $\mathcal{DB}$ combines the sample and the reference and $\mathcal{AS}$ does not receive the reference $\mathcal{E}_{\text{GM}}(b_i, pk)$ but the combination of the sample $\mathcal{E}_{\text{GM}}(b'_i \oplus b_i, pk)$. Using the previous attack, however, $\mathcal{AS}$ can still learn $b'_i$ and the $b'_i \oplus b_i$. Additional measures have to be taken to prevent this, e.g., $\mathcal{DB}$ could be required to sign $\mathcal{E}_{\text{GM}}(b'_i \oplus b_i, pk)$, which will be verified by $\mathcal{M}$. Note that in the case where $\mathcal{AS}$ and $\mathcal{DB}$ collude, these countermeasures are not sufficient anymore.

### B. Barbosa et al. ACISP 2008

*1) Description:* In [20] Barbosa *et al.* presented a new protocol for biometric authentication, following [19] (see previous Section III-A). A notable difference between these two comes from the fact that [19] compares two biometric templates by their Hamming distance, enabling biometric authentication, whereas [20] classifies one biometric template into different classes thanks to the SVM classifier (support vector machine, see [29] for details) leading to biometric identification. Biometric templates are represented as features vector where each feature is an integer, i.e., $b_i = \langle b_{i,1}, \ldots, b_{i,k} \rangle \in \mathbb{N}^k$. Barbosa *et al.* encrypt this vector, feature by feature, with the Paillier cryptosystem [30]. In particular, they exploit its homomorphism property to compute its SVM classifier (think of a sum of scalar products) in the encrypted domain.

However, as we explain further below in this section, as the features are encrypted one by one, an adversary can do something similar as the attack described in the previous section (Section III-A).

Let $\mathcal{E}_{\text{Paillier}}$ (resp. $\mathcal{D}_{\text{Paillier}}$) denote the encryption (resp. decryption) with Paillier's cryptosystem. This cryptosystem enjoys a homomorphic property which ensures that the product of two encrypted texts corresponds to the encryption of their sum: for $m_1, m_2 \in \mathbb{Z}_n$ we have that $\mathcal{D}_{\text{Paillier}}(\mathcal{E}_{\text{Paillier}}(m_1) \times \mathcal{E}_{\text{Paillier}}(m_2)) = m_1 + m_2 \mod n$. Note that $\mathbb{Z}_n$ is the plaintext space of the Paillier cryptosystem.

The SVM classifier takes as input $U$ classes (or users) and $S$ samples per class, and determines support vectors $SV_{i,j}$ and weights $\alpha_{i,j}$ for $1 \leq i \leq S$ and $1 \leq j \leq U$. Following the notation in [20], let $v = (v_1, \ldots, v_k) = b_i$ denote a freshly captured biometric sample. For this sample

the classifier computes

$$cl_{\text{SVM}}^{(j)}(v) = \sum_{i=1}^{S} \alpha_{i,j} \sum_{l=1}^{k} v_l (SV_{i,j})_l \text{ for } j = 1, \ldots, U. \quad (7)$$

With this vector $cl_{\text{SVM}}(v)$, it is possible to determine which class is the most likely for $v$ or to reject it. The support vectors $SV_{i,j}$ and the weight coefficients $\alpha_{i,j}$ are the references that are stored by $\mathcal{DB}$.

Briefly, the scheme of Barbosa *et al.* works as follows:

1) The sensor $\mathcal{S}$ captures a fresh biometric sample and encrypts each of the features of its template $v = (v_1, \ldots, v_k)$ with Paillier's cryptosystem and sends it to the authentication server $\mathcal{AS}$. Let auth = $(\mathcal{E}_{\text{Paillier}}(v_1), \ldots, \mathcal{E}_{\text{Paillier}}(v_k))$.
2) The database $\mathcal{DB}$ computes an encrypted version of the SVM classifier for this biometric data: $c_j = \prod_{i=1}^{S} (\prod_{l=1}^{k} [auth_j]_l^{[SV_{i,j}]_l})^{\alpha_{i,j}}$ where $[.]_l$ denotes the $l^{\text{th}}$ component of a tuple. This $c_j$ corresponds to the encryption of the $cl_{\text{SVM}}^{(j)}$ with Paillier's cryptosystem as defined above. The database returns the values $c_j$ to $\mathcal{AS}$.
3) The authentication server $\mathcal{AS}$ scrambles the values $c_j$ and forwards them to $\mathcal{M}$.[1]
4) The matcher $\mathcal{M}$, using the private key of the system, decrypts the components of the SVM classifier and performs the classification of $v$. The classification returns the class for which the value $cl_{\text{SVM}}^{(j)}$ is maximal.
5) Based on the output of $\mathcal{M}$, $\mathcal{AS}$ determines the real identity of $\mathcal{U}_i$ (in case of non-rejection).

*2) Authentication Server Adversary ($\mathcal{A}=\mathcal{AS}$):* The following attack shows how a malicious $\mathcal{AS}$ can recover the biometric references. In this scheme, the biometric reference data that are stored by $\mathcal{DB}$, i.e., the support vectors $SV_{i,j}$ and the weight coefficients $\alpha_{i,j}$, represent hyperplanes that are used for classification. These $k$-dimensional hyperplanes are expressed as linear combinations of enrolment samples (the support vectors). We will show how these hyperplanes can be recovered dimension by dimension.

Let us rewrite (7) as

$$\begin{aligned} cl_{\text{SVM}}^{(j)}(v) &= v_1 \sum_{i=1}^{S} \alpha_{i,j}(SV_{i,j})_1 + \cdots + v_k \sum_{i=1}^{S} \alpha_{i,j}(SV_{i,j})_k \\ &= v_1 \beta_{j,1} + \cdots + v_k \beta_{j,k}. \end{aligned}$$

By sending a vector $v = \langle 1, 0, \ldots, 0 \rangle$ to $\mathcal{DB}$, $\mathcal{AS}$ will retrieve the encryption of $\beta_{j,1} = \sum_{i=1}^{S} \alpha_{i,j}(SV_{i,j})_1$ for each user, indexed by $j$, in the database.

Instead of sending all $c_j = \mathcal{E}_{\text{Paillier}}(\beta_{j,1})$ to $\mathcal{M}$, only one value will be kept by $\mathcal{AS}$, e.g., $c_1 = \mathcal{E}_{\text{Paillier}}(\beta_{1,1})$. The authentication server will set $c_2 = \mathcal{E}_{\text{Paillier}}(x)$ for some value $x \in \mathbb{Z}_n$ and all other $c_j = \mathcal{E}_{\text{Paillier}}(0)$. The matcher $\mathcal{M}$ will return the index of the class with the greatest value, which is 1 if $\beta_{1,1} > x$ and 2 if $\beta_{1,1} < x$.

The initial value of $x = n/2$. If $\beta_{1,1} > x$ then $\mathcal{AS}$ will adjust $x$ to $n/2 + n/4$, otherwise $x = n/2 - n/4$. By repeating this process and adjusting the value $x$, the exact value $\beta_{1,1}$ can be learned after $\log_2 n$ queries. Hence, the reference data of a single user can be learned in $k \log_2 n$ queries to the matcher.

with the permutation). Quite logical, as the matcher is determining a list of candidates. In particular, although the identifiers are permuted, he can detect if related inputs are used, to trace the user whole database (with a known input)

*3) Discussion:* As in Section III-A this attack succeeds because features are encrypted separately and there is no check to see if the sample and the reference were really merged. The same attack can in principle be used to learn any information about the input sample.

### C. Stoianov SPIE 2010

*1) Description:* In [22], Stoianov introduces several authentication schemes relying on the Blum-Blum-Shub pseudo-random generator. We focus on the database setting from the paper (cf. Section 7 of [22]). In this setting there is a service provider SP that performs the verification. Consistent with our model, we will call this entity the matcher $\mathcal{M}$. Sample and reference are combined before being sent to $\mathcal{M}$ and although this is not explicitly mentioned in [22] we designate this functionality to the authentication server $\mathcal{AS}$ in our model.

In the schemes of [22], the biometric data $b$ are binarized and are combined with a random codeword $c$ coming from an error-correcting code to form a secure sketch or code offset $b \oplus c$ where $\oplus$ stands for the exclusive-or (XOR). When a new capture $b'$ is made, whenever $b'$ is close to $b$ (using the Hamming distance) it is possible to recover $c$ from $b \oplus b' \oplus c$ using error correction. This technique is known as the fuzzy commitment scheme of Juels and Wattenberg [5]. An additional layer of protection is added by encrypting the secure sketch using Blum-Goldwasser.

The Blum-Blum-Shub pseudo-random generator [31] is a tool used in the Blum-Goldwasser asymmetric cryptosystem [32]. From a seed $x_0$ and a public key, a pseudo-random sequence $S$ is generated. In the following, $S$ is XOR-ed to the biometric data to be encrypted. By doing so, the state of the pseudo-random generator is updated to $x_{t+1}$. From $x_{t+1}$ and the private key, the sequence $S$ can be recomputed.

In this system of Stoianov, $\mathcal{M}$ generates the keys and sends the public key to $\mathcal{S}$. On enrollment

1) Sensor $\mathcal{S}$ computes $(S \oplus b \oplus c, x_{t+1})$ where:
   - Sample $b$ is the freshly captured biometric data,
   - String $S$ is a pseudo-random sequence and $x_{t+1}$ is the state of the Blum-Blum-Shub pseudo-random generator as described above, and
   - $c$ is a random codeword which makes the secure sketch $c \oplus b$;
2) Sensor $\mathcal{S}$ sends $S \oplus b \oplus c$ to $\mathcal{DB}$;
3) Sensor $\mathcal{S}$ sends $x_{t+1}$ and $H(c)$ to $\mathcal{M}$ where $H$ is a cryptographic hash function.

Using the private key, $\mathcal{M}$ computes $S$ from $x_{t+1}$ and stores it along $H(c)$. Periodically, $\mathcal{M}$ (resp. $\mathcal{DB}$) updates $S$ (resp. $S \oplus b \oplus c$) to $\check{S}$ (resp. $\check{S} \oplus c \oplus b$) with an independent stream cipher.

During authentication sensor $\mathcal{S}$ receives a new sample $b'$ and forwards $(S' \oplus b', x'_{t+1})$ to $\mathcal{AS}$, where $S'$ is a new pseudo-random sequence. It is assumed that there is some sort of authentication server $\mathcal{AS}$ that retrieves $\check{S} \oplus c \oplus b$ from $\mathcal{DB}$ and merges it with $S' \oplus b'$. Finally $S' \oplus b' \oplus \check{S} \oplus b \oplus c$ and $x'_{t+1}$ are sent to $\mathcal{M}$. Using the private key $\mathcal{M}$ recovers $S'$. From $S'$ and $\check{S}$, $\mathcal{M}$ computes $c \oplus b \oplus b'$, tries to decode it and verifies the consistency of the result with $H(c)$.

---

[1] In [20], the entity that makes the decision is refered to as the verification server. To be consistent with our model we continue to use the term matcher.

*2) Matcher Adversary ($\mathcal{A}=\mathcal{M}$):* Let $\mathcal{M}$ be the primary attacker. It is inherent to the scheme that $\mathcal{M}$ can always trace a valid user over different queries by looking at the codeword $c$, which is revealed after a successful authentication. Depending on the entity that colludes with $\mathcal{M}$ additional attacks can be deviced.

If $\mathcal{M}$ and $\mathcal{DB}$ collude ($\mathcal{A}=\mathcal{M}+\mathcal{DB}$) they learn the sketch $c \oplus b$. This implies that they can immediately trace users with different identities following the linkability attack based on the decoding of the sum of two sketches as described in [11]. From a genuine match, $\mathcal{M}$ learns $c$ and thus also $b$.

If $\mathcal{M}$ and $\mathcal{S}$ collude ($\mathcal{A}=\mathcal{M}+\mathcal{S}$) they control and always learn the input sample $b'$. By setting $b' = 0$ they learn $c \oplus b$ from a single query. If a successful authentication occurred, the adversary learns everything.

If $\mathcal{M}$ and $\mathcal{AS}$ collude ($\mathcal{A}=\mathcal{M}+\mathcal{AS}$) they always learn the input sample $b'$. They can learn the sketch $c \oplus b$ for any reference and thus trace users with different identities as in the case ($\mathcal{A}=\mathcal{M}+\mathcal{DB}$). They learn the reference $b$ after successful authentication.

exhaustive search block by block in case of an accept to reconstruct b + b'...

*3) Authentication Server Adversary ($\mathcal{A}=\mathcal{AS}$):* In the current scheme, bits are not encrypted bit per bit independently. Moreover, they are masked with streams generated via Blum-Blum-Shub and a codeword so attacks as in Sections III-A and III-B are no longer possible. Nevertheless, there is still a binary structure that $\mathcal{AS}$ may exploit.

Assume that $\mathcal{AS}$ knows $S' \oplus b'$ that leads to a positive decision, i.e., $\mathcal{M}$ accepts $b'$ because $\text{d}(b, b') \leq t$. Then $\mathcal{AS}$ can start from $S' \oplus b'$ and add progressively some errors until he reaches a negative result. Then, he backtracks one step by decreasing the error weight by one to come back to the last positive result. This gives $\mathcal{AS}$ an encrypted template $S' \oplus b''$. Consider now the vector $S' \oplus b'' \oplus \check{S} \oplus c \oplus b$ and replace the first bits (say of small length $l$) by a $l$ bits vector $x$.

- For all possible values of $x$, $\mathcal{AS}$ sends the resulting vector (the first block is changed by the value $x$) to $\mathcal{M}$ who acts as a decision oracle.
- If several values give a positive result, then $\mathcal{AS}$ increases the errors on all but the first block.
- This is repeated until only one value of $x$ gives a positive result.
- When this step is reached, $\mathcal{AS}$ has found the value $x$ with no errors, i.e., he learns the first block of $S' \oplus \check{S} \oplus c$.
- $\mathcal{AS}$ proceeds to the next block.

Following this strategy, it is feasible to recover all the bits of $b \oplus b'$. If $\mathcal{AS}$ colludes with $\mathcal{S}$, he can retrieve the full reference template $b$ as soon as $\mathcal{S}$ knows one sample that is close to $b$. We call this attack a center search (cf. Section IV below).

*4) Discussion:* In a way similar to the inherent traceability of users by $\mathcal{M}$, there are no mechanisms described that protect against the database tracing users over different queries, i.e., by tracking $\check{S} + c + b$ lookups.

We note that the matcher $\mathcal{M}$ is very powerful because he knows the secret key, which allows computing $S'$, and $\check{S}$. As soon as $\mathcal{M}$ colludes with one of the other entities he is able to learn everything from a genuine match or a false accept.

### D. Other Schemes

Due to the generic design of our model, several other schemes in the literature fit our model. Nevertheless, as they are not always designed with the same entities, an adaptation might be required. Some others are not compatible at all; for instance those for which the security relies on a user-secret key stored on the user side. We now present a brief overview of the schemes [23], [24] when analyzed in our model.

*ACM MMSec 2010 eSketch:* This scheme of Failla *et al.* is described in [23] following a client-server model. The client corresponds to $\mathcal{AS}$ and the server can be logically separated into $\mathcal{DB}$ and $\mathcal{M}$. The goal of the scheme is to provide anonymous identification. The $\mathcal{DB}$ stores data derived from the biometric reference, in particular secure sketches, and part of the data is encrypted via the Paillier cryptosystem with the corresponding secret key owned by $\mathcal{AS}$. The identification query is implemented through different exchanges between the entities and at one step the same randomness is used to mask all the different reference templates and the masked values are sent to $\mathcal{AS}$. Consequently, an authentication server adversary ($\mathcal{A} = \mathcal{AS}$) learns the whole database after one successful authentication, because the client ($\mathcal{AS}$ in our model) knows the Paillier secret keys. If the adversary consists of the database and the matcher ($\mathcal{A} = \mathcal{DB} + \mathcal{M}$), it is also possible to learn the reference template, which is supposed to be hidden for the server.

*ACM MMSec 2010 Secure Multiparty Computation:* This scheme of Raimondo *et al.* [24] is also described following a client-server model with secure multiparty computation between them to achieve an identification scenario (authentication scenario as well, cf. [24, Fig. 3]). The scheme is not made to be resistant against malicious adversaries. Fitting it in our model, we have $\mathcal{AS}$ which obtains the result and $\mathcal{DB}$ which stores all the references in clear; $\mathcal{AS}$ sends an encrypted (via Paillier) query to $\mathcal{DB}$; $\mathcal{DB}$ sends back to $\mathcal{AS}$ all the entries combined with the query (this gives in fact a database containing all the Euclidean distances) and then $\mathcal{AS}$ and $\mathcal{M}$ interact (secure multiparty protocol) to output the list of identifiers for which the distance is below a threshold. Here again encryption of the query is made block by block, so a similar strategy as in Section III-B is possible when $\mathcal{A} = \mathcal{AS}$. An adversary $\mathcal{A} = \mathcal{DB} + \mathcal{M}$ can also tamper the inputs to the last part of the protocol to learn information about the query.

## IV. FORMALIZATION OF ATTACK SCENARIOS

The goal of this section is to explore some generic attack scenarios that can be used for analyzing actual protocol specifications. These attacks are presented in the framework as described in Section II and generalize the attacks of the previous Section III. As explained in Section II we only consider malicious internal attackers, i.e., $\mathcal{AS}$, $\mathcal{DB}$, $\mathcal{M}$ and combinations of these entities. User $\mathcal{U}_i$ and the sensor $\mathcal{S}$ have been excluded as individual attackers.

### A. Blackbox Attack Model

The different attacks that can be carried out by the attackers are modeled as *blackbox attacks*, following recent results from [25]. This allows us to clearly specify the focus of the attack. Our blackbox-attack model consists of two logical entities:

1) The attacker, i.e., one or more system entities. These entities are fully under control of the attacker: internal data are known, messages can be modified and additional transactions can be generated.
2) The target or the blackbox, i.e., the combination of all other system entities. The attack is focused on the data that are protected by the system components within the blackbox.

The target is modeled as a blackbox because the attacker can only observe the input-output behavior of the box. This adequately reflects remote protocols where only the communication can be seen by the attacker. No details are known about the internal state of the remote components. During the attack, the attacker will "tweak" inputs to the blackbox. However, all communication must comply with the protocol specification. Any messages that are malformed or that are sent in the wrong order are rejected by the blackbox.

It should be noted that there are cases in which the attacker cannot generate additional transactions because he has to follow the protocol specifications. E.g., if $\mathcal{DB}$ is attacking he has to wait until a request is received from $\mathcal{AS}$. When analyzing protocols it should be assumed that this will occur with a reasonable frequency. If relevant, attack complexity can be expressed in function of this frequency. Similarly, if the attacker is $\mathcal{AS}$, he receives inputs from $\mathcal{S}$ and communicates with $\mathcal{DB}$ and $\mathcal{M}$. In this case we exclude $\mathcal{U}_i$ and $\mathcal{S}$ from the blackbox. It should be assumed, though, that a number of inputs from $\mathcal{S}$ is available to $\mathcal{AS}$. This does not necessarily imply that $\mathcal{S}$ is under control of $\mathcal{AS}$. The analysis of the attack can take into account the amount of data that is available.

We will now consider a number of possible adversaries and blackbox attacks in our framework.

### B. Attacker = $\mathcal{AS}$

*Decomposed Reference Attack:* Let's assume that only one reference $b_i$ is returned by $\mathcal{DB}$. The goal of this attack is to learn $b_i$. Biometric samples or references are often represented as a "string", i.e., a concatenation (let $\|$ denote concatenation) of (binary) symbols. Let's assume that $f_2(b_i)$ is the concatenation of a subfunction $\hat{f}_2$ that is applied on each of the $n$ components $b'_{i,j}$ of $b'_i$ individually. If $\mathcal{AS}$ has to combine $f_2(b_i)$ and $f_1(b'_i)$ without knowing either the sample or the reference, it is likely that $f_1$ and $f_3$ will also be the concatenation of component-wise applied subfunctions, i.e., $f_3(b_i, b'_i) = \hat{f}_3(b_{i,1}, b'_{i,1}) \| \ldots \| \hat{f}_3(b_{i,n}, b'_{i,n})$. Note that in our model $\mathcal{AS}$ can generate the value $\hat{f}_3(b_{i,j}, b'_{i,j})$ but this value should not reveal to $\mathcal{AS}$ whether the inputs are the same or not. This decomposition of references is used in the scheme analyzed in Section III-A and the following attack applies to it.

Suppose that $\mathcal{AS}$ is able to generate a value that is valid output of $\hat{f}_3$ when the two component inputs $b_{i,j}$ and $b'_{i,j}$ are the same and similarly when they are not the same, e.g., the output is the encryption of one or zero. If $\mathcal{AS}$ can also compute $f_1$, then $\mathcal{AS}$ can fully reconstruct $b_i$. To do so $\mathcal{AS}$ choose the first component of $b'_i$ at random, combines it with the first component of $b_i$ and sends the result to $\mathcal{M}$. The other components that are sent to $\mathcal{M}$ are such that $t$ of those are an output of $\hat{f}_3$ that reflects different inputs and the $n-t-1$ remaining components are outputs that reflect equal inputs. Note that $t$ is the comparison threshold. If the guess of $\mathcal{AS}$ for the first component is correct then $\mathcal{M}$ will return a positive match. Otherwise the guess is wrong and $\mathcal{AS}$ can try again. This process can be repeated until all components of $b_i$ are recovered. For binary samples, this requires $n$ queries to $\mathcal{M}$ and 1 query to $\mathcal{DB}$.

As shown in Section III-B a similar attack can be executed if the biometric data are represented as real-valued or integer-valued feature vectors. However, more queries might be required to get an accurate result.

*Center Search Attack Using $\mathcal{S}$:* In this attack, $\mathcal{S}$ is also compromised and under the control of the attacker. The attack goal is to learn the full reference $b_i$ from a close sample. The input sample is obviously always known to $\mathcal{AS}$ and $\mathcal{S}$. Thus at some point in time $\mathcal{U}_i$ will present a sample $b'_i$ that matches reference $b_i$. This sample will lie at some distance from the reference. In the case where biometrics are represented as binary strings and the system implements a hamming distance matcher the attacker can recover the exact $b_i$ as follows.

The sensor flips the first bit of $b'_i$ and sends the new sample to $\mathcal{AS}$ who performs the whole authentication procedure. If the authentication succeeds $\mathcal{S}$ flips the second bit, leaving the first bit also flipped, and sends the sample to $\mathcal{AS}$ who follows the procedure again. This continues until the sample no longer matches $b_i$. Then the sensor starts again by restoring the first bit of the sample that is no longer accepted and forwards it to $\mathcal{AS}$. If it gets accepted this means that the first bit of the original sample $b'_i$ was the same as the first bit of $b_i$. If not, then the first bits were different. One by one the bits in $b'_i$ that are different from those in $b_i$ can be corrected. This technique was demonstrated in Section III-C.

We call this the center search attack because we start from a sample that lies in a sphere with radius $t$, the matching threshold, and the reference as center point. The goal of this attack is to move the sample to the center of the sphere. The worst-case complexity of this attack for bitstrings of length $n$ is the greatest of $2*t+n$ and $4t$. The complexity is $2*t+n$ if there are $t-1$ bit-errors in the beginning and one at the end of the string. The first $t-1$ errors get corrected by flipping them and $t$ additional bits need to be flipped to invalidate the sample. Locating the bit-errors requires searching till the end of the string where the last error is. The complexity is $4t$ if there are $t-1$ correct bits followed by $t$ wrong bits. So $2t$ flips are needed before the queries no longer match and then

$2t$ positions need to be searched. In practice, $t \leq n/2$ and thus the worst-case complexity is $2t + n$.

*False-Acceptance-Rate Attack:* A false acceptance occurs if a sample, not coming from $\mathcal{U}_i$, is close enough to $b_i$ to be recognized by the system as a sample coming from $\mathcal{U}_i$. The name comes from the fact that an attacker can take a large existing database of samples and feed that to a biometric authentication system. Due to the inherent false acceptance rates, there will be a sample in the attacker's database that matches the reference in the system with high probability.

The goal of this attack is to learn $b_i$ from a matching sample that is unknown to the attacker. This attack combines ideas from the previous attacks. The attacker is $\mathcal{AS}$, not including $\mathcal{S}$, and $\mathcal{AS}$ does not know how to compute an output of $f_3$ that reflects equal (or different) inputs. It is assumed, however, that the attacker can replace the components of $b'_i$ in the value he received from $\mathcal{S}$, i.e., $f_1(b'_i)$. This is definitely the case if $f_1$ is a concatenation of subfunctions and if $\mathcal{AS}$ can compute such subfunction $\hat{f}_1$.

The actual attack then proceeds as follow. The attacker $\mathcal{AS}$ waits until a genuine user presents a valid sample. The attack is similar as in the center-search attack, only now $\mathcal{AS}$ will not flip bits but simply replace them with a known value, e.g., one. He will do this until the sample no longer matches. Then $\mathcal{AS}$ already knows that the last bit he replaced was not one and he will restore that bit. Then he continues to substitute the bits one by, carefully observing whether the sample matches or not and learning all the bits. The first bits that were flip to invalidate the sample can be learned by simply restoring them.

### C. Attacker = $\mathcal{DB}$ or $\mathcal{M}$

The attacker is the database $\mathcal{DB}$ or the matcher $\mathcal{M}$ who communicate with the authentication service $\mathcal{AS}$ only. The attackers cannot achieve any of the attack goals individually because their blackboxes give output, which cannot be influenced by the attacker, before receiving input. If these entities do not collude with other entities they are simply passive attackers and by Assumption 2 they cannot mount any attacks.

*Including the sensor $\mathcal{S}$:* If the sensor colludes with the database or the matcher, some attacks are trivial: the provided sample is known and thus it is also easy to trace a user based on the provided sample or identity.

A powerful attacker is the combination of the $\mathcal{M}$ and the $\mathcal{S}$, as was shown in Section III-C. Because the sensor can send any input and any identity, the attacker does not have to wait for a matching sample. The same center-search attack can be performed as in the case where $\mathcal{AS}$ and $\mathcal{S}$ are the attacker.

### D. Attacker = $\mathcal{AS}$ and $\mathcal{DB}$

The attackers ($\mathcal{AS}$ and $\mathcal{DB}$) receive fresh input from $\mathcal{S}$ and communicate with the matcher. They can search the entire database and turn to identification although the protocol could be designed to operate in verification mode.

The input sample $b'_i$ can be learned in the same way as the $b_i$ was learned in the attacks of $\mathcal{AS}$, if the same conditions hold. Then, depending on the implementation, the attacker can learn the entire database because $\mathcal{DB}$ will return any $b_i$ and $\mathcal{AS}$ will manipulate it until all bits are known.

### E. Attacker = $\mathcal{AS}$ and $\mathcal{M}$

The attack goal with the highest impact is to learn the reference $b_i$ from the database. Depending on how the $\mathcal{M}$ implements its functionality this can be a very powerful attacker, e.g., if $\mathcal{M}$ possesses decryption keys for encrypted samples/templates as was the case in the schemes analyzed in Sections III-A, III-B and III-C.

### F. Attacker = $\mathcal{DB}$ and $\mathcal{M}$

In this combination of attackers, $\mathcal{DB}$ will manipulate its output so that it can be of use to the $\mathcal{M}$. The relevant attack goals are to learn $b'_i$ and to trace $U_i$.

### G. Attacker = $\mathcal{AS}$ and $\mathcal{DB}$ and $\mathcal{M}$

In this particular case, the attacker is a combination of $\mathcal{AS}$, $\mathcal{DB}$ and $\mathcal{M}$, and the goal is to learn $b'_i$. If the reference $b_i$ is not stored in the clear by the database, the attacker may want to learn $b_i$ also. Tracing $\mathcal{U}_i$ is almost trivial because the attacker can perform a search (identification) on the database $\mathcal{DB}$. The attack goals are easily reached if the data can be decomposed as explained in the attacks of $\mathcal{AS}$.

## V. CONCLUSION

Biometric authentication protocols that are found in the literature are usually designed in the honest-but-curious model assuming that there are no malicious insider adversaries. In this paper, we have challenged that assumption and shown how some existing protocols are not secure against such adversaries. Such analysis is extremely relevant in the context of independent database providers. Much attention was given to an authentication server attacker, which is a central and powerful entity in our model. To prevent the attacks that were presented, stronger enforcement of the protocol design is needed: many attacks succeed because transactions can be duplicated or manipulated.